# Negentropy concept revisited: Standard thermodynamic properties of 16 bacteria, fungi and algae species

Marko Popovic

University of Belgrade, Belgrade, 11000, Serbia

**Abstract:** Standard molar and specific (per gram) enthalpy of formation, entropy and Gibbs free energy of formation of biomatter have been determined for 16 microorganism species, including Methylococcus capsulatus, Klebsiella aerogenes, Paracoccus denitrificans, Escherichia coli, Pseudomonas C12B, Aerobacter aerogenes, Magnetospirillum gryphiswaldense, Saccharomyces cerevisiae, Candida utilis, Chlorella, Chlorella a sp. MP-1, C. minutissima, C. pyrenoidosa and C. vulgaris. The average values of $\Delta_f H^o$ are for bacteria -4.22 kJ/g, for fungi -5.03 kJ/g and for algae -4.40 kJ/g. The average values of $S^o$ are for bacteria 1.48 J/g K, for fungi 1.45 J/g K and for algae 1.48 J/g K. The average values of $\Delta_f G^o$ are for bacteria -2.30 kJ/g, for fungi -3.15 kJ/g and for algae -2.48 kJ/g. Based on the results, an analysis was made of colony growth in time. In the first three phases, entropy change of microorganisms plated in a Petri dish is positive and entropy of the colony increases. In the fourth phase, due to limitation of nutrients, entropy remains constant. In the fifth phase, due to lack of nutrients, entropy of the colony decreases and microorganisms die-off. Based on the results, the negentropy concept is analyzed.

**Keywords:** Microorganisms; Standard specific entropy; Standard specific enthalpy of formation; Standard specific Gibbs free energy of formation; Growth.

## 1. Introduction

Animate matter represents a highly organized, self-assembled amount of substance clearly separated by a semipermeable membrane from its environment (inanimate matter) [Morowitz, 1992]. In biological terms, an organism is characterized by cellular structure that corresponds to a thermodynamic system [Morowitz, 1955; Schrödinger, 2003; von Bertalanffy, 1950; Balmer, 2011; Popovic, 2017a, 2018, Demirel, 2014]. Animate matter represents an open system [von Bertalanffy, 1950]. Moreover, because of the seven characteristics of life [Campbell and Reece, 2002], it also represents a growing open system due to accumulation of matter and energy [Popovic, 2017b]. Thus, animate matter performs thermodynamic processes corresponding to biological (life) processes.

In the late of the 19th century, Boltzmann was the first to attempt to relate life processes to the fundamental laws of nature, using entropy: *"The general struggle for existence of animate beings is not a struggle for raw materials, but a struggle for entropy, which becomes available through the transition of energy from the hot sun to the cold earth"* [Boltzmann, 1974]. Boltzmann's reasoning was expanded by Schrödinger [2003]: an organism *"feeds upon negative entropy, attracting, as it were, a stream of negative entropy upon itself, to compensate the entropy increase it produces by living and thus to maintain itself on a stationary and fairly low entropy level."* Thus, thermodynamic entropy of a cell or an organism is predicted to decrease during its life span [Balmer, 2011; Davies et al., 2013]. Balmer [2011] argued that *"one characteristic that seems to make a living system unique is its peculiar affinity for self-organization"*. As the system lives, it grows and ages and generally becomes more complex. So, *"living systems are uniquely characterized by decreasing their entropy over their life spans"* [Balmer, 2011]. However, the decrease in entropy of organisms during life seems to be controversial because all matter has positive entropy according to the third law. Growth represents accumulation of matter. Due to accumulation, entropy change during life should be positive.

Schrödinger [2003] pointed out a potential conflict between the second law and life processes, because direction of change in entropy tends to its maximum according to the second law, and the direction of change in life process seemed to be toward greater order, decreasing thermodynamic entropy and accumulating information. Schrödinger explained this "apparent" contradiction by suggesting that the very existence of living systems depends on increasing the entropy of their surroundings [2003]. The second law is not violated but only locally circumvented at the expense of global increase in thermodynamic entropy, concluded Morowitz [1968]. Therefore, based on Boltzmann's reasoning that entropy is a measure of disorder, Schrödinger [2003] introduced a quantitative measure of order – negentropy. Negentropy was proposed as entropy taken with a negative sign: -(entropy).

The second law of thermodynamics is the "entropy law" and represents a law of disorder, a view due to Boltzmann "*Because there are so many more possible disordered states than ordered ones, a system will almost always be found either in the state of maximum disorder*" [1974]. Disorder and order in living organisms were considered by Schrödinger [2003], who argued:

> "*Life seems to be orderly and lawful behavior of matter, not based exclusively on its tendency to go over from order to disorder, but based partly on existing order that is kept up… If D is a measure of disorder, its reciprocal, 1/D, can be regarded as a direct measure of order. Since the logarithm of 1/D is just minus the logarithm of D, we can write Boltzmann's equation thus:*
>
> $$-(entropy) = k \log(1/D)$$
>
> (where *k* is the Boltzmann constant and *log* is the natural logarithm). *Hence the awkward expression negative entropy can be replaced by a better one: entropy, taken with the negative sign, is itself a measure of order.*"

Schrödinger postulated a local decrease in entropy for living systems, quantifying it with negentropy and explaining it through order of biological structures. Order (1/*D*) represents the number of states that cannot arrange randomly, exemplified by replication of genetic code.

Negentropy has found applications in biothermodynamics - thermodynamics of living organisms, describing their orderliness and explaining the general patterns of metabolism. However, recent studies are casting doubt on the validity of negentropy [Popovic, 2017a; Annamalai and Nanda, 2017; Annamalai and Silva, 2012; Silva and Annamalai, 2008].

The goal of this paper is to examine the negentropy concept, by determining the change in entropy in various microorganism species and comparing the results with the predictions of the negentropy concept. Section 2 discusses the methods used for determining biomatter thermodynamic parameters. Section 3 gives the results and discusses general trends. Finally, section 4 applies the results to growth of *Escherichia Coli* colony, calculating the entropy of an *E. coli* cell made of biomatter and water, determining entropy change during growth of the entire bacterial colony, and discussing the negentropy concept.

Negentropy Concept Revisited

## 2. Methods

Empirical formulas have been collected from the literature for 16 microorganism species, including 8 bacteria, 2 yeast and 6 algae species. They are given in tables 1 and 2. From tables 1 and 2, it can be seen that the most abundant elements in all microorganisms are C, H, O and N, which can be used to roughly represent their composition. Thus, the average chemical formula of bacteria was found from data in table 1 to be $CH_{1.8}O_{0.4}N_{0.2}$, for fungi it is $CH_{1.8}O_{0.5}N_{0.2}$ and for algae $CH_{1.8}O_{0.5}N_{0.1}$. All other elements are present in amounts a magnitude lower or less.

A living organism consists of biomatter and water. Composition of microorganism biomatter can be used to calculate its entropy, according to Battley [1999], using the empirical equation

$$S^o_{mol,bio} = 0.187 \sum S^o_{mol,atoms} \qquad (1)$$

where $S^o_{mol,bio}$ is standard molar entropy of biomatter in a microorganism. $S^o_{atoms}$ is sum of the standard entropies of individual elements in an empirical formula of the biomass, weighed by their respective coefficients. Battley's [1999] empirical formula predicts the composition of dried biomass. The effect of neglecting entropy of hydration is the greatest source of error and leads to an error not greater than 19.7% [Battley, 1999]. Similarly, standard molar entropy of formation $\Delta_f S^o_{mol,bio}$ of microorganism biomatter from elements can be calculated from its composition [Battley, 1999], using the equation

$$\Delta_f S^o_{mol,bio} = -0.813 \sum S^o_{mol,atoms} \qquad (2)$$

Similarly to entropy, standard molar enthalpy of formation of microorganisms can be calculated from their elemental composition using Thornton's rule [Battley, 1999, 1998]. Thornton's rule states that enthalpy of combustion $\Delta_c H$ of organic matter is proportional to the number equivalents in the organic matter $E$.

$$\Delta_c H = -111.14 \; kJ \; eq^{-1} \cdot E \qquad (3)$$

The number of equivalents in an organic substance is the number of electrons transferred to oxygen during combustion to $CO_2(g)$, $H_2O(l)$, $N_2(g)$, $P_4O_{10}(s)$ and $SO_3(g)$ [Battley, 1998]. Inorganic ions, like $Na^+$ and $Mg^{2+}$ are not included, since they are already in their highest oxidation state and cannot transfer any electrons to oxygen [Battley, 1998]. Thus, $E$ is calculated through the equation

$$E = 4\, n_C + n_H - 2\, n_O - 0\, n_N + 5\, n_P + 6\, n_S \qquad (4)$$

where $n_C$, $n_H$, $n_O$, $n_N$, $n_P$ and $n_S$ are the number of C, H, O, N, P and S atoms in the empirical formula of a microorganism [Battley, 1998]. If any of these atoms are not present, they are just neglected during calculation [Battley, 1998].

Once $\Delta_c H$ is determined, standard molar enthalpy of formation of microorganism biomatter $\Delta_f H^o_{mol,bio}$ can be calculated as enthalpy of the first reactant in the oxidation reaction [Battley, 1998]

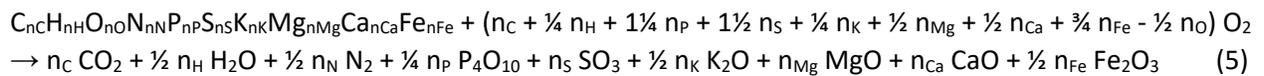

$C_{nC}H_{nH}O_{nO}N_{nN}P_{nP}S_{nS}K_{nK}Mg_{nMg}Ca_{nCa}Fe_{nFe}$ + ($n_C$ + ¼ $n_H$ + 1¼ $n_P$ + 1½ $n_S$ + ¼ $n_K$ + ½ $n_{Mg}$ + ½ $n_{Ca}$ + ¾ $n_{Fe}$ - ½ $n_O$) $O_2$
→ $n_C$ $CO_2$ + ½ $n_H$ $H_2O$ + ½ $n_N$ $N_2$ + ¼ $n_P$ $P_4O_{10}$ + $n_S$ $SO_3$ + ½ $n_K$ $K_2O$ + $n_{Mg}$ $MgO$ + $n_{Ca}$ $CaO$ + ½ $n_{Fe}$ $Fe_2O_3$ \qquad (5)

that is, using the formula

$$\Delta_f H^o_{mol,bio} = n_C \Delta_f H^o_{mol,CO2} + \frac{1}{2} n_H \Delta_f H^o_{mol,H2O} + \frac{1}{4} n_P \Delta_f H^o_{mol,P4O10} + n_S \Delta_f H^o_{mol,SO3} +$$
$$\frac{1}{2} n_K \Delta_f H^o_{mol,K2O} + n_{Mg} \Delta_f H^o_{mol,MgO} + n_{Ca} \Delta_f H^o_{mol,CaO} + \frac{1}{2} n_{Fe} \Delta_f H^o_{mol,Fe2O3} - \Delta_c H \quad (6)$$

As with entropy, the greatest source of error in determination of $\Delta_f H°_{mol,bio}$ is enthalpy of hydration, which leads to an error of 2% [Battley, 1998].

From $\Delta_f H°_{mol,bio}$ and $\Delta_f S°_{mol,bio}$, it is possible to calculate the standard molar Gibbs free energy of formation of microorganisms from elements $\Delta_f G°_{mol,bio}$, using the Gibbs equation

$$\Delta_f G^o_{mol,bio} = \Delta_f H^o_{mol,bio} - T \Delta_f S^o_{mol,bio} \quad (7)$$

where under standard conditions $T$ = 298.15 K. The error in $\Delta_f G°_{mol,bio}$, $\sigma_G$ can be estimated from error in $\Delta_f H°_{mol,bio}$, $\sigma_H$, and $\Delta_f S°_{mol,bio}$, $\sigma_S$, using the equation $\sigma_G = \sigma_H + T \sigma_S$.

### 3. Results

Standard molar thermodynamic parameters of biomatter have been calculated, as described in section 2, including enthalpy of formation from elements $\Delta_f H°_{mol}$, entropy $S°_{mol}$ and Gibbs free energy of formation from elements $\Delta_f G°_{mol}$. They are given in table 1. From the standard molar thermodynamic properties, it is possible to calculate standard specific (per gram) thermodynamic properties using the equation $X°_g = X°_{mol}/M_r$, where $X°_g$ is standard property ($\Delta_f H$, $S$ or $\Delta_f G$) per gram, $X°_{mol}$ is the corresponding property per mole and $M_r$ is molar weight of the biomatter. Thus, standard specific enthalpy of formation from elements $\Delta_f H°_g$, specific entropy $S°_g$ and specific Gibbs free energy of formation from elements $\Delta_f G°_g$ of biomatter were calculated and are given in table 2.

The thermodynamic parameters have very similar values for the three classes, as would be expected. The average values of $\Delta_f H°_g$ are for bacteria -4.22 kJ/g, for fungi -5.03 kJ/g and for algae -4.40 kJ/g. The average values of $S°_g$ are for bacteria 1.48 J/g K, for fungi 1.45 J/g K and for algae 1.48 J/g K. The average values of $\Delta_f G°_g$ are for bacteria -2.30 kJ/g, for fungi -3.15 kJ/g and for algae -2.48 kJ/g.

Interesting trends can be seen among the standard thermodynamic properties. For all the studied microorganisms, $\Delta_f H°_g < 0$, $S°_g > 0$ and $\Delta_f G°_g < 0$. The $\Delta_f H°_g < 0$ trend means that formation of biomatter from elements is exothermic. The $S°_g > 0$ trend is a consequence of the third law of thermodynamics, stating that entropy cannot have a negative value. Finally, the $\Delta_f G°_g < 0$ implies that formation of biomatter from elements is a spontaneous process for all the studied microorganism species.

### 4. Discussion

In this section, the results from section 3 will be applied to bacterial colony growth. Entropy of a single *E. coli* cell will be calculated, entropy change of a growing *E. coli* colony will be analyzed, and the negentropy concept discussed. The following analysis is based on two axioms:

1. All microorganisms grow and multiply, making colonies, through cell division.
2. Growth is caused by import and accumulation of substances taken from the surroundings, resulting in change in mass and volume of a colony.

Starting from axioms 1 and 2, it can be concluded that mass of a microorganism colony increases during its lifespan, increasing with each cell division: d$m$/d$t$ > 0, where $m$ is colony mass and $t$ is time.

Growth of microorganisms occurs in five phases, during time and culture aging: (1) lag, (2) exponential, (3) declining growth rate, (4) stationary and (5) death phase. Duration of the phases depend on availability of nutrients in the environment. The exponential phase is the period when growth is the most intense. The number of microorganisms is described as a function of time by the equation

$$N_{cells} = N_{cells,0}\ 2^{t/t_d} \tag{8}$$

where $N_{cells,0}$ is the initial cell number and $t_d$ is division time [Widdel, 2007]. For *Escherichia coli,* the division time is $t_d$ = 40 min [Neidhardt, 1996]. A cell or a colony, as growing open systems, are characterized by thermodynamic parameters, including entropy [Battley, 1999; von Stockar and Liu, 1999].

Entropy of a single *E. coli* cell was calculated using data from table 2. A cell consists of biomatter and water. Therefore, the total entropy of a cell, $S_{cell}$, is

$$S_{cell} = S^o_{g,bio} \cdot m_{bio} + S^o_{g,w} \cdot m_w + \Delta_{hyd}S \tag{9}$$

where $S^o_{g,bio}$ is standard specific entropy of biomatter (table 2), $m_{bio}$ mass of biomatter, $m_w$ mass of water in the cell, and $S^o_{g,w}$ standard specific entropy of water, which is 3.886 J/g K [Cox *et al.*, 1984]. Equation (9) also contains entropy of hydration of biomatter, $\Delta_{hyd}S$. However, there is no method to accurately predict $\Delta_{hyd}S$. Thus, $\Delta_{hyd}S$ was included into the error of $S^o_{g,bio}$ given in table 2, as is described in section 2, and $\Delta_{hyd}S$ was set to zero in equation 9. Finally, a single *Escherichia coli* cell weighs $9.5 \cdot 10^{-13}$ g, containing $2.8 \cdot 10^{-13}$ g (30%) of dry mass and $6.7 \cdot 10^{-13}$ g (70%) of water [Neidhardt, 1996]. Therefore, according to equation (9) the entropy of a single *E. coli* cell is $(3.01 \pm 0.05) \cdot 10^{-12}$ J/K.

Based on entropy of a single cell and colony growth rate, entropy of an *E. coli* colony was calculated as a function of time. Microorganisms live in colonies. The entropy of a colony is the sum of entropies of all microorganisms that comprise it

$$S_{colony} = N_{cells} \cdot S_{cell} \tag{10}$$

Since a colony grows during time through increase in $N_{cells}$, its entropy changes. Therefore, using equations (8), (9) and (10), entropy of a microorganism colony can be determined as a function of time, as is shown for *E. coli* in Figure 1.

As can be seen from Figure 1, there is an exponential increase in both the number of cells and in entropy of the colony during the exponential phase of colony growth. This is a general trend, according to equation (10), since standard specific entropy of any substance, including living organisms, can only be positive due to the third law of thermodynamics, as discussed in section 3. Therefore, any growing organism increases its mass and entropy.

The change in cell number, mass and entropy of a microorganism colony is positive during the first three phases, equal to zero during the fourth and negative during the fifth phase. Based on this, two stages can be distinguished in the existence of a microorganism colony: (1) accumulation phase (life) when cell number, mass and entropy increase, and (2) decumulation phase (death) when they decrease. If an *E. coli* colony begins growth with $10^6$ cells and an initial entropy of $3.01 \cdot 10^{-6}$ J/K, then through import, accumulation of matter from the environment and cell division, the colony grows, and increases its mass and entropy to 101.03 J/K, after 1000 min. It is obvious that entropy of microorganisms, during colony life and aging, increases during the first three phases. During the fourth phase, cell number and entropy remain constant in time. Only in the last phase, due to lack of nutrients, the entropy of the colony begins to decrease due to loss of living cells. So, the existence of a microorganism colony can be divided into two phases: accumulation phase when the number of cells and entropy increase, and decumulation phase when the number of cells and entropy decrease. Finally, once the last cell has decomposed, the colony ceases to exist.

The results given in tables 1 and 2 show that entropy during growth and accumulation of substances increases. The increase in entropy of the colony (or an organism) represents a problem for the negentropy concept. Also, Gibbs energy change is negative during growth, implying the spontaneity of the process.

**5. Conclusions**

Based on elemental compositions (empirical formulas) of 16 microorganism species, including 8 bacteria, 2 yeast and 6 algae species, standard molar and specific enthalpies of formation from elements $\Delta_f H^o$, entropies $S^o$ and Gibbs free energies of formation from elements $\Delta_f G^o$ of biomatter have been calculated (tables 1 and 2). For all the analyzed species, formation from elements is exothermic $\Delta_f H^o < 0$ and spontaneous $\Delta_f G^o < 0$, while entropy of biomatter is positive $S^o > 0$ in accordance with the third law of thermodynamics.

Bacteria, fungi and algae colonies begin their lifespan from a certain number of cells, which are characterized by a certain mass, volume and entropy. During the colony lifespan, microorganisms take nutrients from their environment, accumulating them and performing growth. This process leads to accumulation of mass, and increase in cell number and entropy. Entropy of a growing (accumulating) bacterial colony increases, as shown in figure 1. The increase in entropy of the colony (or an organism) represents a problem for the negentropy concept.

The existence of a microorganism colony can be divided into two stages: accumulation phase (life) when the number of cells and entropy increase and decumulation phase (death) when the number of cells and entropy decrease. Finally, once the last cell has decomposed, the colony ceases to exist.

**Table 1:** Standard molar enthalpy of formation $\Delta_f H°_{mol,bio}$, entropy $S°_{mol,bio}$, and Gibbs free energy of formation $\Delta_f G°_{mol,bio}$ of biomatter (all matter except water) for various bacteria, fungi and algae. Empirical formulas reflect elemental composition of microorganisms and are normalized per mole of carbon atoms. To appreciate the size of a cell, a single *Magnetospirillum gryphiswaldense* cell can be described by $C_{2.31\times10^{10}}H_{4.68\times10^{10}}O_{2.93\times10^{9}}N_{6.56\times10^{1}}Fe_{4.00\times10^{7}}$. The thermodynamic parameters for entire cells are given by the equation $X°_{cell}= X°_{mol,bio}\cdot n_{bio} + X°_{mol,w}\cdot n_w$ where $X°_{cell}$ is thermodynamic parameter ($\Delta_f H°$, $S°$ or $\Delta_f G°$) for a single cell, $X°_{mol,bio}$ thermodynamic parameter for biomatter from this table, $n_{bio}$ number of moles of dry matter, $X°_{mol,w}$ thermodynamic parameter for water and $n_w$ number of moles of water in the cell (Section 4).

| Name | Formula | Reference | $\Delta_f H°_{mol}$ (kJ/mol) | $S°_{mol}$ (J/mol K) | $\Delta_f G°_{mol}$ (kJ/mol) |
|---|---|---|---|---|---|
| *Bacteria* | | | | | |
| Bacteria | $CH_{1.666}O_{0.270}N_{0.200}$ | Abbott and Clamen, 1973 | -61.90 ± 1.24 | 30.15 ± 5.94 | -22.82 ± 3.01 |
| *Methylococcus capsulatus* | $CH_{2.000}O_{0.500}N_{0.270}$ | van Dijken and Harder, 1975 | -123.64 ± 2.47 | 39.90 ± 7.86 | -71.93 ± 4.82 |
| Klebsiella aerogenes | $CH_{1.750}O_{0.430}N_{0.220}$ | Naresh et al., 2011 | -100.14 ± 2.00 | 34.60 ± 6.82 | -55.28 ± 4.04 |
| K. aerogenes | $CH_{1.730}O_{0.430}N_{0.240}$ | Naresh et al., 2011 | -99.50 ± 1.99 | 34.72 ± 6.84 | -54.50 ± 4.03 |
| K. aerogenes | $CH_{1.750}O_{0.470}N_{0.170}$ | Naresh et al., 2011 | -109.03 ± 2.18 | 34.47 ± 6.79 | -64.34 ± 4.21 |
| K. aerogenes | $CH_{1.730}O_{0.430}N_{0.240}$ | Naresh et al., 2011 | -99.50 ± 1.99 | 34.72 ± 6.84 | -54.50 ± 4.03 |
| Paracoccus denitrificans | $CH_{1.810}O_{0.510}N_{0.200}$ | Stouthamer, 1977 | -119.83 ± 2.40 | 36.51 ± 7.19 | -72.50 ± 4.54 |
| P. denitrificans | $CH_{1.510}O_{0.460}N_{0.190}$ | Shimizu et al., 1978 | -99.18 ± 1.98 | 31.71 ± 6.25 | -58.08 ± 3.85 |
| Escherichia coli | $CH_{1.770}O_{0.490}N_{0.240}$ | Bauer and Ziv, 1976 | -114.11 ± 2.28 | 36.36 ± 7.16 | -66.98 ± 4.42 |
| Pseudomonas C12B | $CH_{2.000}O_{0.520}N_{0.230}$ | Mayberry et al., 1968 | -128.09 ± 2.56 | 39.56 ± 7.79 | -76.80 ± 4.89 |
| Aerobacter aerogenes | $CH_{1.830}O_{0.550}N_{0.250}$ | Naresh et al., 2011 | -129.35 ± 2.59 | 38.42 ± 7.57 | -79.55 ± 4.84 |
| Magnetospirillum gryphiswaldense | $CH_{2.060}O_{0.130}N_{0.280}Fe_{0.00174}$ | Naresh et al., 2011 | -44.02 ± 0.88 | 33.72 ± 6.64 | -0.31 ± 2.86 |
| *Fungi* | | | | | |
| S. cerevisiae | $CH_{1.613}O_{0.557}N_{0.158}P_{0.012}S_{0.003}K_{0.022}Mg_{0.003}Ca_{0.001}$ | Battley, 1999 | -131.99 ± 2.64 | 34.66 ± 6.83 | -87.07 ± 4.68 |
| Saccharomyces cerevisiae | $CH_{1.640}O_{0.520}N_{0.160}$ | Harrison, 1967 | -116.65 ± 2.33 | 33.91 ± 6.68 | -72.69 ± 4.32 |

| | | | | | | | | | | |
|---|---|---|---|---|---|---|---|---|---|---|
| S. cerevisiae | $CH_{1.830}O_{0.560}N_{0.170}$ | Kok and Roels, 1980 | -131.58 | ± | 2.63 | 37.18 | ± | 7.32 | -83.38 | ± | 4.82 |
| S. cerevisiae | $CH_{1.810}O_{0.510}N_{0.170}$ | Wang et al, 1976 | -119.83 | ± | 2.40 | 35.97 | ± | 7.09 | -73.19 | ± | 4.51 |
| Candida utilis | $CH_{1.830}O_{0.540}N_{0.100}$ | Herbert, 1976 | -127.13 | ± | 2.54 | 35.54 | ± | 7.00 | -81.06 | ± | 4.63 |
| C. utilis | $CH_{1.870}O_{0.560}N_{0.200}$ | Naresh et al., 2011 | -132.85 | ± | 2.66 | 38.20 | ± | 7.53 | -83.32 | ± | 4.90 |
| C. utilis | $CH_{1.830}O_{0.460}N_{0.190}$ | Naresh et al., 2011 | -109.35 | ± | 2.19 | 35.62 | ± | 7.02 | -63.18 | ± | 4.28 |
| C. utilis | $CH_{1.870}O_{0.560}N_{0.200}$ | Naresh et al., 2011 | -132.85 | ± | 2.66 | 38.20 | ± | 7.53 | -83.32 | ± | 4.90 |
| *Algae* | | | | | | | | | | | |
| Algae | $CH_{2.481}O_{1.038}N_{0.151}P_{0.00943}$ | Wang et al., 2017 | -260.31 | ± | 5.21 | 54.03 | ± | 10.64 | -190.27 | ± | 8.38 |
| Chlorella (algae) | $CH_{1.719}O_{0.404}N_{0.175}P_{0.0105}$ | Manahan and Manahan, 2009 | -95.34 | ± | 1.91 | 33.00 | ± | 6.50 | -52.56 | ± | 3.85 |
| Chlorella a sp. MP-1 | $CH_{1.793}O_{0.608}N_{0.121}$ | Phukan et al., 2011 | -141.17 | ± | 2.82 | 36.78 | ± | 7.25 | -93.49 | ± | 4.98 |
| C. minutissima | $CH_{1.714}O_{0.286}N_{0.143}$ | Prajapati et al., 2014 | -66.93 | ± | 1.34 | 30.02 | ± | 5.91 | -28.02 | ± | 3.10 |
| C. pyrenoidosa | $CH_{1.625}O_{0.250}N_{0.125}$ | Prajapati et al., 2014 | -56.15 | ± | 1.12 | 27.92 | ± | 5.50 | -19.96 | ± | 2.76 |
| C. vulgaris | $CH_{1.667}O_{0.222}N_{0.111}$ | Prajapati et al., 2014 | -51.30 | ± | 1.03 | 27.65 | ± | 5.45 | -15.47 | ± | 2.65 |

**Table 2:** Standard specific (per gram) enthalpy of formation $\Delta_f H°_{g,bio}$, entropy $S°_{g,bio}$, and Gibbs free energy of formation $\Delta_f G°_{g,bio}$ of biomatter (all matter except water) for various bacteria, fungi and algae. Empirical formulas reflect elemental composition of microorganisms and are normalized per mole of carbon atoms. To appreciate the size of a cell, a single *Magnetospirillum gryphiswaldense* cell can be described by $C_{2.31\times10^{10}}H_{4.68\times10^{10}}O_{2.93\times10^{9}}N_{6.56\times10^{1}}Fe_{4.00\times10^{7}}$. The thermodynamic parameters for entire cells are given by the equation $X°_{cell} = X°_{g,bio} \cdot m_{bio} + X°_{g,w} \cdot m_w$ where $X°_{cell}$ is thermodynamic parameter ($\Delta_f H°$, $S°$ or $\Delta_f G°$) for a single cell, $X°_{g,bio}$ thermodynamic parameter for biomatter from this table, $m_{bio}$ mass of the cell dry matter, $X°_{g,w}$ thermodynamic parameter for water and $m_w$ mass of water in the cell (Section 4).

| Name | Formula | Reference | $\Delta_f H°_{g,bio}$ (kJ/g) | $S°_{g,bio}$ (J/g K) | $\Delta_f G°_{g,bio}$ (kJ/g) |
|---|---|---|---|---|---|
| *Bacteria* | | | | | |
| Bacteria | $CH_{1.666}O_{0.270}N_{0.200}$ | Abbott and Clamen, 1973 | -2.97 ± 0.06 | 1.45 ± 0.29 | -1.10 ± 0.14 |
| *Methylococcus capsulatus* | $CH_{2.000}O_{0.500}N_{0.270}$ | van Dijken and Harder, 1975 | -4.79 ± 0.10 | 1.55 ± 0.30 | -2.79 ± 0.19 |
| *Klebsiella aerogenes* | $CH_{1.750}O_{0.430}N_{0.220}$ | Naresh et al., 2011 | -4.22 ± 0.08 | 1.46 ± 0.29 | -2.33 ± 0.17 |
| *K. aerogenes* | $CH_{1.730}O_{0.430}N_{0.240}$ | Naresh et al., 2011 | -4.15 ± 0.08 | 1.45 ± 0.29 | -2.27 ± 0.17 |
| *K. aerogenes* | $CH_{1.750}O_{0.470}N_{0.170}$ | Naresh et al., 2011 | -4.61 ± 0.09 | 1.46 ± 0.29 | -2.72 ± 0.18 |
| *K. aerogenes* | $CH_{1.730}O_{0.430}N_{0.240}$ | Naresh et al., 2011 | -4.15 ± 0.08 | 1.45 ± 0.29 | -2.27 ± 0.17 |
| *Paracoccus denitrificans* | $CH_{1.810}O_{0.510}N_{0.200}$ | Stouthamer, 1977 | -4.83 ± 0.10 | 1.47 ± 0.29 | -2.92 ± 0.18 |
| *P. denitrificans* | $CH_{1.510}O_{0.460}N_{0.190}$ | Shimizu et al., 1978 | -4.21 ± 0.08 | 1.35 ± 0.27 | -2.47 ± 0.16 |
| *Escherichia coli* | $CH_{1.770}O_{0.490}N_{0.240}$ | Bauer and Ziv, 1976 | -4.57 ± 0.09 | 1.45 ± 0.29 | -2.68 ± 0.18 |
| *Pseudomonas C12B* | $CH_{2.000}O_{0.520}N_{0.230}$ | Mayberry et al., 1968 | -5.01 ± 0.10 | 1.55 ± 0.30 | -3.00 ± 0.19 |
| *Aerobacter aerogenes* | $CH_{1.830}O_{0.550}N_{0.250}$ | Naresh et al., 2011 | -4.95 ± 0.10 | 1.47 ± 0.29 | -3.04 ± 0.19 |
| *Magnetospirillum gryphiswaldense* | $CH_{2.060}O_{0.130}N_{0.280}Fe_{0.00174}$ | Naresh et al., 2011 | -2.18 ± 0.04 | 1.67 ± 0.33 | -0.02 ± 0.14 |
| *Fungi* | | | | | |
| *Saccharomyces cerevisiae* | $CH_{1.613}O_{0.557}N_{0.158}P_{0.012}S_{0.003}K_{0.022}Mg_{0.003}Ca_{0.001}$ | Battley, 1999 | -5.04 ± 0.10 | 1.32 ± 0.26 | -3.32 ± 0.18 |
| *S. cerevisiae* | $CH_{1.640}O_{0.520}N_{0.160}$ | Harrison, 1967 | -4.82 ± 0.10 | 1.40 ± 0.28 | -3.00 ± 0.18 |
| *S. cerevisiae* | $CH_{1.830}O_{0.560}N_{0.170}$ | Kok and Roels, 1980 | -5.22 ± 0.10 | 1.48 ± 0.29 | -3.31 ± 0.19 |

Negentropy Concept Revisited| | | | | | | | | | |
|---|---|---|---|---|---|---|---|---|---|
| S. cerevisiae | $CH_{1.810}O_{0.510}N_{0.170}$ | Wang et al, 1976 | -4.92 | ± | 0.10 | 1.48 | ± | 0.29 | -3.00 ± 0.18 |
| Candida utilis | $CH_{1.830}O_{0.540}N_{0.100}$ | Herbert, 1976 | -5.32 | ± | 0.11 | 1.49 | ± | 0.29 | -3.39 ± 0.19 |
| C. utilis | $CH_{1.870}O_{0.560}N_{0.200}$ | Naresh et al., 2011 | -5.18 | ± | 0.10 | 1.49 | ± | 0.29 | -3.25 ± 0.19 |
| C. utilis | $CH_{1.830}O_{0.460}N_{0.190}$ | Naresh et al., 2011 | -4.58 | ± | 0.09 | 1.49 | ± | 0.29 | -2.65 ± 0.18 |
| C. utilis | $CH_{1.870}O_{0.560}N_{0.200}$ | Naresh et al., 2011 | -5.18 | ± | 0.10 | 1.49 | ± | 0.29 | -3.25 ± 0.19 |
| *Algae* | | | | | | | | | |
| Algae | $CH_{2.481}O_{1.038}N_{0.151}P_{0.00943}$ | Wang et al., 2017 | -7.77 | ± | 0.16 | 1.61 | ± | 0.32 | -5.68 ± 0.25 |
| Chlorella (algae) | $CH_{1.719}O_{0.404}N_{0.175}P_{0.0105}$ | Manahan and Manahan, 2009 | -4.15 | ± | 0.08 | 1.44 | ± | 0.28 | -2.29 ± 0.17 |
| Chlorella a sp. MP-1 | $CH_{1.793}O_{0.608}N_{0.121}$ | Phukan et al., 2011 | -5.59 | ± | 0.11 | 1.46 | ± | 0.29 | -3.70 ± 0.20 |
| C. minutissima | $CH_{1.714}O_{0.286}N_{0.143}$ | Prajapati et al., 2014 | -3.30 | ± | 0.07 | 1.48 | ± | 0.29 | -1.38 ± 0.15 |
| C. pyrenoidosa | $CH_{1.625}O_{0.250}N_{0.125}$ | Prajapati et al., 2014 | -2.89 | ± | 0.06 | 1.44 | ± | 0.28 | -1.03 ± 0.14 |
| C. vulgaris | $CH_{1.667}O_{0.222}N_{0.111}$ | Prajapati et al., 2014 | -2.73 | ± | 0.05 | 1.47 | ± | 0.29 | -0.82 ± 0.14 |

**Figure 1:** Thermodynamics of growth of an *Escherichia coli* colony. (a) Number of cells as a function of time determined using equation (8). $N_{cells,0} = 1·10^6$ was used. (b) Entropy as a function of time determined using equation (10). For *E. coli* it was found in section 4 that $S_{cell} = (3.01 ± 0.05) · 10^{-12}$ J/K.

(a)

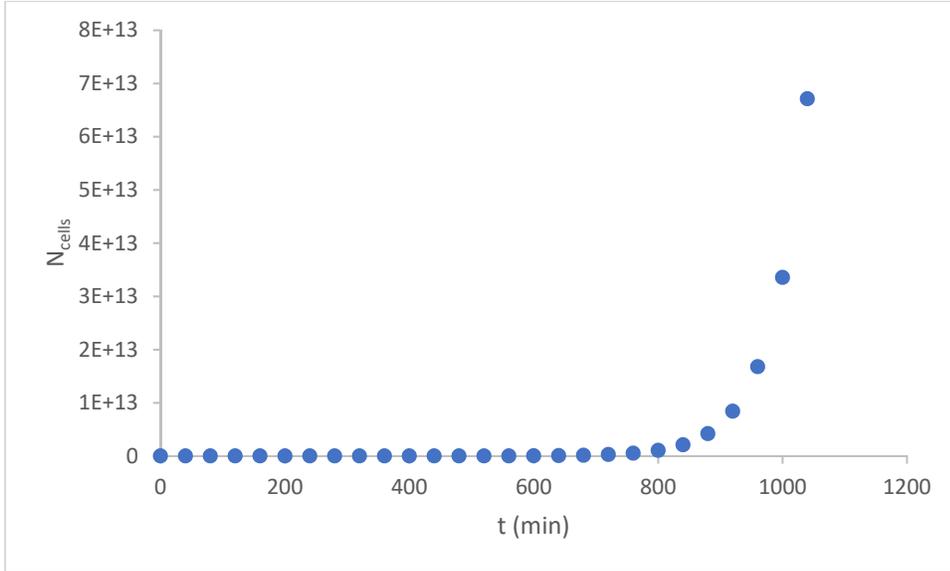

(b)

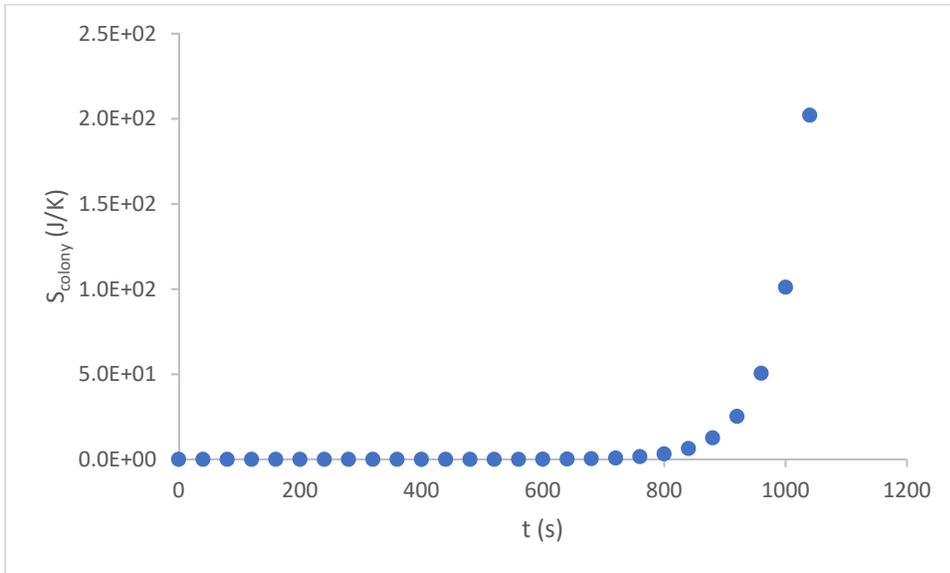